# Reducing the hydrogen content in liquid helium

Dominik Sifrig[a], Sascha Martin[a], Dominik Zumbühl[a], Christian Schönenberger[a], Laurent Marot[a]*

[a]Department of Physics, University of Basel, Klingelbergstrasse 82, 4056 Basel, Switzerland

*Corresponding author: tel.: +41 61 207 37 20, e-mail: laurent.marot@unibas.ch (L. Marot)

**Abstract**

Helium has the lowest boiling point of any element in nature at normal atmospheric pressure. Therefore, any unwanted substance like impurities present in liquid helium will be frozen and will be in solid form. Even if these solid impurities can be easily eliminated by filtering, liquid helium may contain a non-negligible quantity of molecular hydrogen. These traces of molecular hydrogen are the causes of a known problem worldwide: the blocking of fine-capillary tubes used as flow impedances in helium evaporation cryostats to achieve temperatures below 4.2 K. This problem seriously affects a wide range of cryogenic equipment used in low-temperature physics research and leads to a dramatic loss of time and costs due to the high price of helium. Here, we present first the measurement of molecular hydrogen content in helium gas. Three measures to decrease this molecular hydrogen are afterward proposed; (i) improving the helium quality, (ii) release of helium gas in the atmosphere during purge time for the regeneration cycle of the helium liquefier's internal purifier, and (iii) installation of two catalytic converters in a closed helium circuit. These actions have eliminated our low-temperature impedance blockage occurrences now for more than two years.

1. Introduction

Even though helium (He) is the second most common element in the universe, there is only a limited amount available on the earth. Helium resources of the world were estimated to be about 35.2 billion cubic meters [1]. The locations and volumes of the significant deposits, in billion cubic meters, are Qatar, 10.1; Algeria, 8.2; Russia, 6.8; USA, 3.9; Canada, 2.0; and China, 1.1. With the current worldwide helium consumption being around 160 million cubic meters per year [1], the worldwide reserves are estimated to sustain the supply for $\approx$ 220 years at the present consumption rates. Although the long-term prospect is not as severe as the short-term landscape of helium supply, helium remains an unrecoverable natural resource. Unless recycled, any use cuts into the limited reserves and pushes us closer to the days when it is used up. As demand from high-tech manufacturing, including demand from China, is also rising, the problem is, helium is being used up faster than it can be produced these days. That created chaos in the helium supply chain in 2018, as roughly 30 percent of the global supply was taken off the market [2]. The price has gone up by 50 to 100 %, depending on local circumstances and position in the supply chain. Therefore, scientific laboratories are seeking to recycle helium. This is precisely the policy of the Department of Physics at the University of Basel and will be detailed later.

Another concern about liquid helium is the hydrogen ($H_2$) content. The blocking of fine-capillary tubes used as flow impedances in He evaporation cryostats to achieve temperatures below 4.2 K is generally attributed to nitrogen or air impurities entering these tubes from the main bath [3]. However, even if one prevents these impurities from entering the capillary tubes, blocking of the capillary tubes frequently occurred. This points to hydrogen as the source of blocking. Many research laboratories around the world have faced this issue at a considerable financial cost because the affected systems have to be warmed up to room temperature to recover their typical low-temperature operation performance [4–8]. In 2016, Gabal et al. [4] and, also in a recent paper, Will et al. [9], it was reported that even hydrogen concentrations within parts per billion are sufficient to block the helium flow within a few hours of operation. On the other hand, hydrogen impurities in liquid helium are easy to detect using a dedicated sensor developed at the University of Zaragoza [4]. It consists of a test capillary that is connected to a vacuum pump (see figure 2 in reference [10]) and has been successfully tested elsewhere [9]. Here, we report on another approach for the quantification hydrogen impurities $H_2$ in liquid helium. In this work, we focus on the determination and reducing the amount of $H_2$ in liquid He. In the following, improvement of helium quality from 4.6 to 5.0, the release of He gas into the atmosphere during the purge time for the regeneration cycle of helium liquefier's internal purifier and the installation of two catalytic converters in our closed helium circuit will be discussed.

## 2. Liquefier

The Department of Physics at the University of Basel is equipped for more than 20 years with a helium liquefier. In September 2016, a new liquefier L70 [11] from Linde Kryotechnik was installed. It has been reported by Berdais et al. [12], its further developments were explained by Clausen et al. [13], and improvement of the redesigned internal purifier was presented by Decker et al. [14]. The liquefaction performance is at about 30 l/h without liquid pre-cooling. To avoid the release of helium into the atmosphere and reduce the depletion of helium reserves in the coming years [1], we have been equipped with a complete helium recovery plant. The recovery plant is schematically presented in Figure 1, as well as in reference [15]. With this installation, the amount of helium gas recovered varied from ca. 78 to 86 %, depending on the year. To address the impedance blockage, we have additionally equipped our plant with a setup for nitrogen ($N_2$) and $H_2$ measurements (Figure 1). The locations, at which $H_2$ and $N_2$ concentration can be measured, were marked in red, labelled 1 to 4.

Hydrogen quantification is carried out using a continuous analyzer (Nova 430LRM Series Low Range Hydrogen Analyzer) [16] in a range of 0-200 parts-per-million (ppm) and with a resolution of 1 ppm. It is operating in a continuous mode for the measurement of low levels of $H_2$ in air, $N_2$, or He samples, containing oxygen. If oxygen is not present, then an air supply system is usually included. $H_2$ is detected using a long-life electrochemical sensor, composed of an anode, cathode, and a suitable electrolyte sealed together. When exposed to hydrogen, the all produces a small output current, corresponding to the concentration of $H_2$ present in the sample. This output is then amplified and directed to the digital display meter. For the measurement of $N_2$ a continuous Binary Gas Analyzer (BGA244 Stanford Research Systems) has been employed.

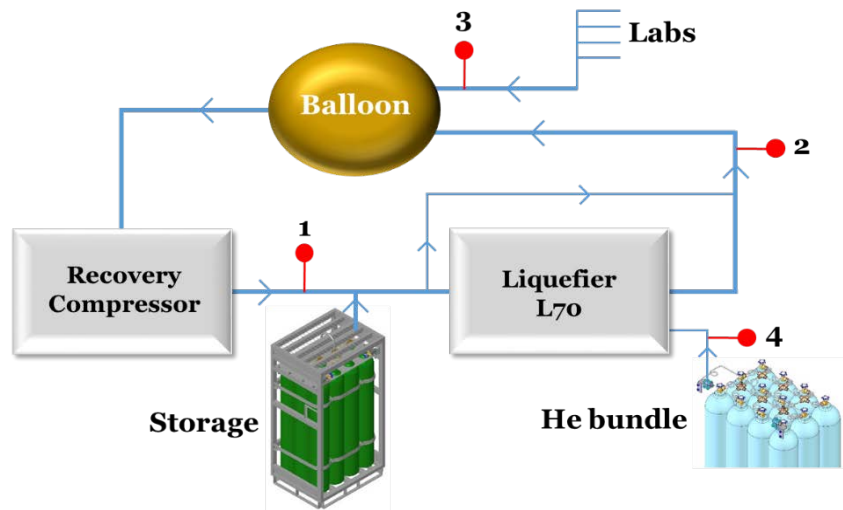

Figure 1: Helium recovery plant including the L70 liquefier, the balloon, the recovery compressor including storage, and the feeding of fresh He delivered in gas bundles. Hydrogen and nitrogen measurements are possible at locations 1 to 4.

### 3. Results

In commercially available helium liquefiers, the internal gas purity is determined by adjusting the cold end temperature, the cold flow, the regeneration completion temperature, and the heater temperature. Typically, the internal purifier operating mode of a helium liquefier starts purification by going through the processes of purge, cool down 1, standby, and cool down 2 and then performs regeneration and enters the purge mode again [15]. As shown in the internal purifier flow diagram in figure 1 of reference [5], the operating conditions of the internal purifier are determined by setting the cold end temperature TI3475, cold flow F3410, regeneration completion temperature TI3465, and heater control output R3470. For Ikeda et al. [5], the operating conditions of a Linde liquefier (L280) were set to a cold end temperature at TI3475 at 32.5 K, cold flow F3410 at 120 l/min, regeneration completion temperature TI3465 at 140 K, and heater control output R3470 at 18 %. In our case, while using a liquid helium liquefier (L70), the parameters were 31 K, 65 l/min, 130 K, and 15 %, respectively. With this configuration, the hydrogen amount in our system measured at location 1 (Figure 1) is plotted in Figure 2. We noticed that during the purge process, the $H_2$ amount was above 200 ppm and was saturating the analyzer. Such purge events are see in the plot as the large signals. Additionally, over a longer time range, a high content of $H_2$ was measured.

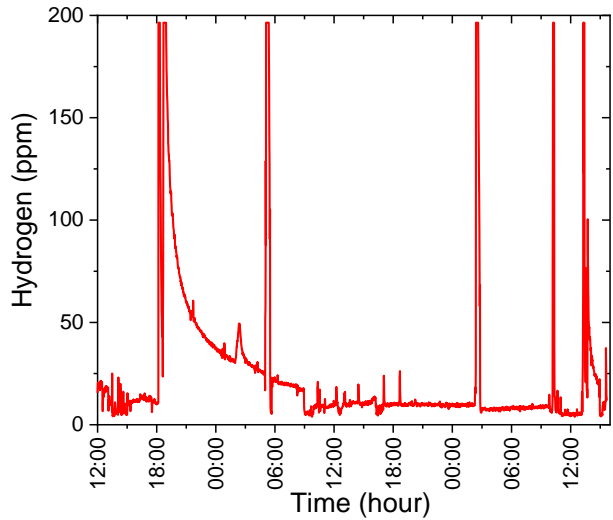

Figure 2: Hydrogen content measured at location 2 of Figure 1.

The blockage of fine-capillary tubes used as flow impedances was first observed at the Department of Physics in 2017. By changing the cold end temperature of the internal purifier (TI3475) from 32.5 to 22 K, Ikeda et al. succeeded in improving the recovered gas purity during purifier regeneration operation from 33.5 to 99 % [5]. The same procedure was then tested with our L70 liquefier. The change from 31 to 26 K (or even lower) of the cold end temperature was not successful in our case. Therefore, a set of measures were taken to address this problem: (i) improving the helium quality from 4.6 to 5.0, (ii) release of helium gas into the atmosphere during the purge time for the regeneration cycle of helium liquefier's internal purifier, and (iii) installation of two catalytic converters in our closed helium circuit.

### 3.1 Change of the helium purity

It has been known for a long time that molecular H is naturally present in helium gas as obtained from natural gas sources [17]. Different methods are typically applied to eliminate it before large-scale helium liquefaction for its distribution worldwide [18,19]. For example, in the case of Palmerston helium plant, in Wickham Point (Australia), the product quality is liquid helium at a 99.999% purity level, containing less than 10 ppm total impurity within a maximum of 1 ppm of $H_2$ [19]. Moreover, ultra-high pure commercial grade He gas, 99.9999 % pure (quality 6.0), containing less than 1 ppm of total impurities in volume, may contain up to 0.5 ppm in volume of $H_2$.

In the case of our recovery plant, less than ca. 20 % of He gas is lost in our closed circuit, new helium bundles are regularly fed into our loop (Figure 1). The quality of the He bundle was changed from 4.6 (purity $\geq$ 99.996 %) to 5.0 (purity $\geq$ 99.999 %) with $H_2$<1 ppm. For the 5.0 quality, oxygen, hydrogen and water are less than 3 ppm. In position 4 (Figure 1) the $H_2$ content was measured for He 4.6 and 5.0 to be in the range of 7-13 ppm and below 1 ppm, respectively. In Figure 3, $H_2$ content measured at location 2 for 4.6 He quality without purge and with 5.0 He quality with a purge is presented. Even if the purge effect has to be taken into account, the total content of $H_2$ has been reduced from 24.3 to 0.8 ml. The concept of this "purge" will be explained in the following section.

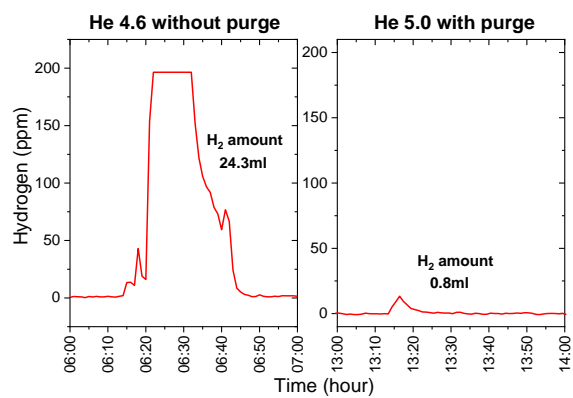

Figure 3: Hydrogen content measured at location 2 of Figure 1 with He 4.6 without purging and He 5.0 with purging.

3.2 Release of helium gas into the atmosphere during the purge time for the regeneration cycle of the internal purifier

As reported by Decker et al., a new 20 K adsorber has been installed downstream of the heat exchanger E3470 for the regeneration cycle of the liquefier's internal purifier (Figure 3 of reference [14]). This small adsorber (A3474) entraps traces of hydrogen and neon up to 100 ppm (in volume) in sum per cycle. For the liquefier purification phase, as specified by Linde company, the gas trapped by the adsorber has been released to the closed circuit. To avoid an increase in the $H_2$ concentration over time, the procedure was modified to implement a release into the atmosphere during the purge time at the end of the regeneration cycle. For this purpose, a new valve (V3410-1) has been installed in parallel to the V3410 (Figure 3 of reference [14]). By adjusting the time of the release to the atmosphere, the maximum $H_2$ level has successfully been lowered by a factor of 10 (Figure 4).

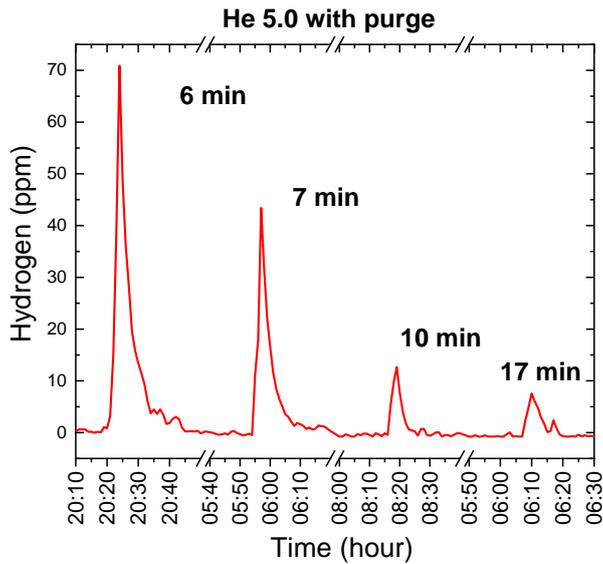

Figure 4: Hydrogen content measured in position 2 of Figure 1 for several purge runs time during the regeneration cycle of the internal purifier.

3.3 Installation of two catalytic converters in the closed helium circuit.

As reported by Haberstroh, car catalytic converter was already implemented by the University of Frankfurt (Germany) in 2013 [6]. Common catalysts are most often a mix of precious metals, mostly from the platinum group and typically palladium [20]. Two catalytic converters were installed in our helium recovery plant to reduce hydrogen. Hydrogen molecules are chemisorbed in a dissociated form onto the catalyst surface and further oxidized to $H_2O$ (g) by the $O_2$ impurities present in the incoming gas. One of the catalytic converters has been installed at the high-pressure part (Hydrogen burner NEO312 neo hydrogen sensors GmbH [21] ) before the dryer shown in figure 3 of reference [13]. The other catalytic converter, a natural gas car exhaust (VW-Touran), was mounted before the recovery compressor (Figure 1). For test purposes, both catalytic converters were mounted directly before the hydrogen measurement analyzer. Using a bottle of nitrogen-containing 100 ppm of $H_2$ and oxygen gas flowing through the catalytic converter, we measured the hydrogen content release for several catalytic converter temperatures. Comparing Figure 5a) and b), we noticed that hydrogen burner NEO312 has to be chemically activated through heat. At 60 C, half of the $H_2$ amount is oxidized, and at 100 C no more $H_2$ is released. The reaction onset temperature is the temperature where a significant fuel consumption begins [22] and, for this particular catalytic converter, is between 30 to 60 C. In comparison, the natural gas car exhaust can already oxidize the entire $H_2$ content at 30 C. Following, the consumption rate of $H_2$ is increased for a temperature of 45 C as compared to 30 C (Figure 5b).

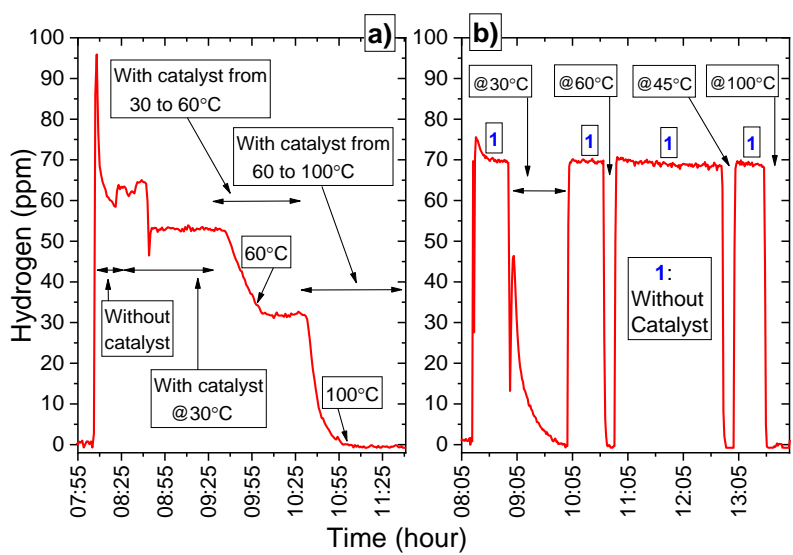

Figure 5: Hydrogen concentration measured using both catalytic converters for several temperatures: a) Hydrogen burner NEO312 and b) natural gas car exhaust (VW-Touran).

### 4. Conclusion

To address the capillary blocking issue that started to occur at the Department of Physics at the University of Basel in 2017 using helium evaporation cryostats, a set of countermeasures were: (i) improving the helium quality from 4.6 to 5.0, (ii) release of helium gas into the atmosphere during the purge time for the regeneration cycle of the helium liquefier's internal purifier and (iii) installation of two catalytic converters in our closed helium circuit. After one year of implementing and testing these three measures, their validity has been continuously demonstrated from the beginning of 2018 by producing hydrogen-free liquid helium. No further blocking of capillary tubes were reported.

### 5. Acknowledgments